\documentclass{article}
\usepackage{frascatiphys_R}
\begin{document}
\title{ 
NONLEPTONIC B DECAYS AND RARE DECAYS
}
\author{
Gerhard Buchalla\\
\em Ludwig-Maximilians-Universit\"at M\"unchen, Sektion Physik, \\
\em Theresienstra\ss e 37, D-80333 M\"unchen, Germany
}
\maketitle
\baselineskip=11.6pt
\begin{abstract}
We review selected topics in the field of nonleptonic
and rare $B$ meson decays. We concentrate in particular
on exclusive channels, discussing recent deve\-lop\-ments
based on the concepts of factorization in QCD and the
heavy-quark limit.
\footnote{
Preprint LMU 05/03}
\end{abstract}
\baselineskip=14pt
\section{Introduction}

The major goal of $B$ physics is to provide us with
novel and decisive tests of the quark flavour sector.
The most interesting $B$ decay channels typically have
small branching fractions below $10^{-4}$ and are being
studied by the current generation of $B$ physics facilities.
Important examples of such decays are nonleptonic modes
as $B\to\pi\pi$ or $B\to\pi K$, and the radiative rare decays
$B\to K^*\gamma$, $\rho\gamma$, $K^* l^+l^-$, $l\nu\gamma$.
They consitute a rich source of information, in particular on
CKM angles and flavour-changing neutral currents (FCNC).
Many new results are becoming available from the $B$ factories.
Both inclusive and exclusive decays can be exploited. Loosely
speaking, the exclusive channels are easier for experiment
while they are harder for theory. The challenge for theory
is to control the effects of QCD. To achieve this it is
necessary to devise a systematic factorization of short-distance
and long-distance contributions, which usually results in
a considerable simplification of the problem.
For $B$ decay matrix elements this factorization relies on the
hierarchy $m_b\gg\Lambda_{QCD}$. This allows us to perform
an expansion around the heavy-quark limit and to factorize
perturbative contributions (scales of order $m_b$)
from nonperturbative dynamics ($\Lambda_{QCD}$).
Since the general concept of factorization in QCD has recently 
found new applications in the important domain of 
{\it exclusive\/} $B$ decays, we shall focus the following
presentation on this area.


\section{Exclusive hadronic $B$ decays in QCD}

The calculation of $B$-decay amplitudes, such as 
$B\to D\pi$, $B\to\pi\pi$ or $B\to \pi K$, starts from
an effective Hamiltonian, which has, schematically, the form
\begin{equation}
{\cal H}_{eff}=\frac{G_F}{\sqrt{2}}\lambda_{CKM}\, C_i Q_i
\end{equation}
Here $C_i$ are the Wilson coefficients at a scale $\mu\sim m_b$,
$Q_i$ are local, dimension-6 operators and $\lambda_{CKM}$ represents
the appropriate CKM matrix elements. The main theoretical problem
is to evaluate the matrix elements of the operators $\langle Q_i\rangle$
between the initial and final hadronic states. A typical matrix element
reads $\langle\pi\pi|(\bar ub)_{V-A}(\bar du)_{V-A}|B\rangle$.

\begin{figure}[t]
\vspace{4cm}
\includegraphics{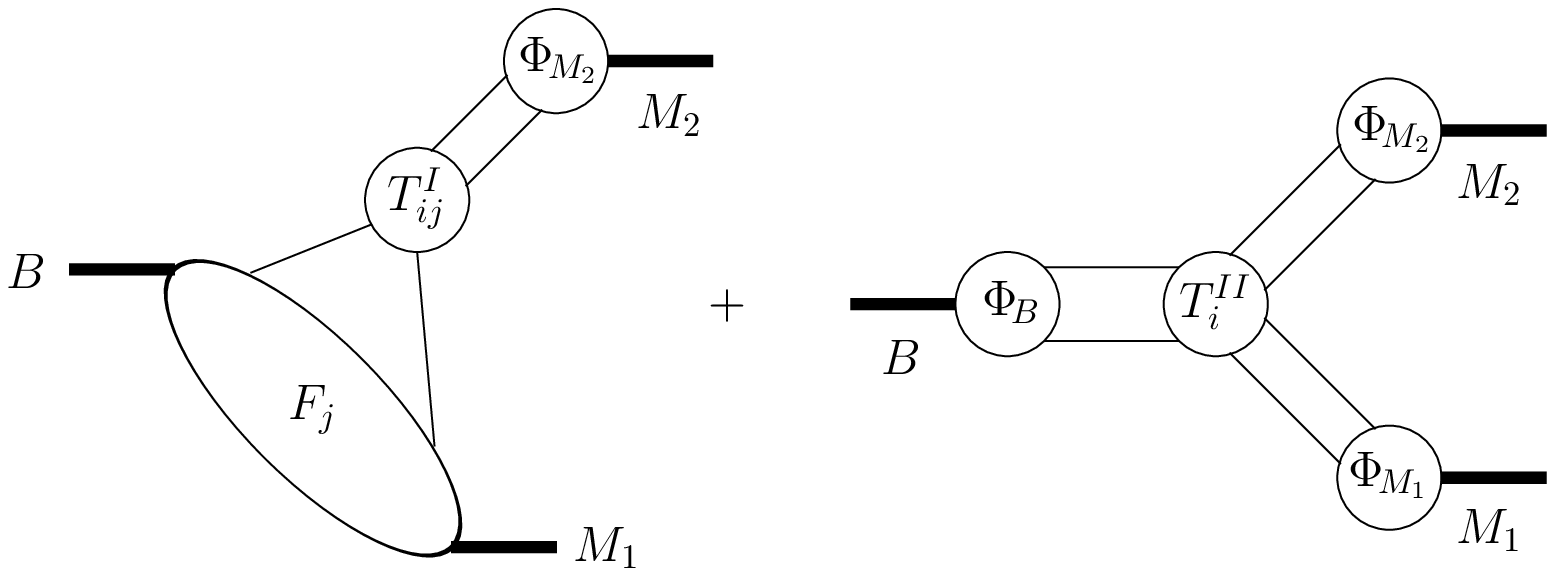}
\caption{\it Graphical representation of the factorization
formula. \label{fig:fform}}
\end{figure}

These matrix elements simplify in the heavy-quark limit, where
they can in general be written as the sum of two terms, each
of which is factorized into hard scattering functions $T^I$ and $T^{II}$,
respectively, and the nonperturbative, but simpler, form factors
$F_j$ and meson light-cone distribution amplitudes $\Phi_M$ 
(Fig. \ref{fig:fform}).
Important elements of this approach are: i) The expansion
in $\Lambda_{QCD}/m_b\ll 1$, consistent power counting, and
the identification of the leading power contribution, for which the
factorized picture can be expected to hold.
ii) Light-cone dynamics, which determines for instance the
properties of the fast light mesons. The latter are described by
light-cone distribution amplitudes $\Phi_\pi$ of their
valence quarks defined as 
\begin{equation}
\langle\pi(p)|u(0)\bar d(z)|0\rangle=\frac{if_\pi}{4}\ \gamma_5\not\! p
\ \int^1_0 dx\ e^{ix pz}\ \Phi_\pi(x)
\end{equation}
with $z$ on the light cone, $z^2=0$.
iii) The collinear quark-antiquark pair dominating the
interactions of the highly energetic pion decouples from soft
gluons (colour transparency). This is the intuitive reason behind
factorization. iv) The factorized amplitude consists of hard, short- 
distance components, and soft, as well as collinear, long-distance
contributions.

\begin{figure}[t]
\vspace{7cm}
\includegraphics{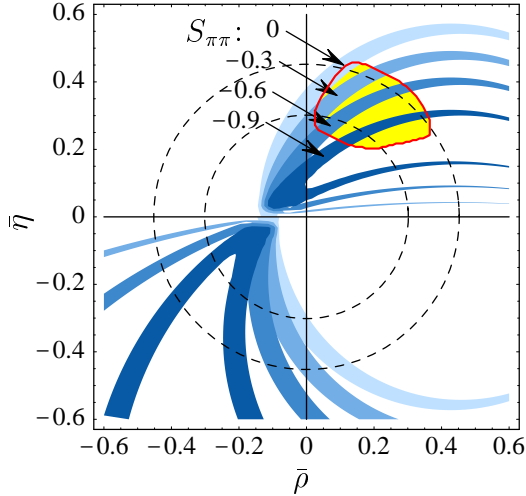}
\caption{\it Constraints in the $\bar\rho$, $\bar\eta$ plane
from CP violation observable $S$ in $B\to\pi^+\pi^-$.
The constraints from $|V_{ub}/V_{cb}|$ (dashed circles)
and from the standard analysis of the unitarity triangle
(irregular shaded area) are also shown. \label{fig:spipi}}
\end{figure}

More details on the factorization formalism can be found
elsewhere \cite{BBNS}. Here we would like to emphasize an
important phenomenological application.
Consider the time-dependent, mixing-induced 
CP asymmetry in $B\to\pi^+\pi^-$
\begin{eqnarray}
{\cal A}_{CP}(t) &=&
\frac{\Gamma(B(t)\to\pi^+\pi^-)-\Gamma(\bar B(t)\to\pi^+\pi^-)}{
      \Gamma(B(t)\to\pi^+\pi^-)+\Gamma(\bar B(t)\to\pi^+\pi^-)}   \\
&=& -S \sin(\Delta M_d t)+ C \cos(\Delta M_d t)
\end{eqnarray}
Using CKM-matrix unitarity, the decay amplitude consists of
two components with different CKM factors and different hadronic
parts, schematically
\begin{equation}
A(B\to\pi^+\pi^-) = V^*_{ub} V_{ud} ({\rm up} - {\rm top})+
  V^*_{cb} V_{cd} ({\rm charm} - {\rm top}) 
\end{equation}
If the penguin contribution $\sim V^*_{cb} V_{cd}$ could be
neglected, one would have $C=0$ and $S=\sin 2\alpha$, hence a direct
relation of ${\cal A}_{CP}$ to the CKM angle $\alpha$.
In reality the penguin contribution is not negligible compared 
to the dominant tree contribution $\sim V^*_{ub} V_{ud}$.
The ratio of penguin and tree amplitude, which enters the
CP asymmetry, depends on hadronic physics.
This complicates the relation of observables $S$ and $C$
to CKM parameters. QCD factorization of $B$-decay matrix elements
allows us to compute the required hadronic input and to determine
the constraint in the ($\bar\rho$, $\bar\eta$) plane implied by
measurements of the CP asymmetry. This is illustrated for $S$
in Fig. \ref{fig:spipi}. 
The widths of the bands indicate the 
theoretical uncertainty \cite{BBNS3}. Note that the constraints
from $S$ are relatively insensitive to theoretical or
experimental uncertainties. The analysis of direct CP violation
measured by $C$ is more complicated due to the importance of
strong phases. The current experimental results are 
\begin{equation}
\begin{array}{cc}
C=-0.94^{+0.31}_{-0.25}\pm 0.09 \quad ({\rm Belle}) & 
-0.30\pm 0.25\pm 0.04 \quad ({\rm Babar})\\
S=-1.21^{+0.38}_{-0.27}{}^{+0.16}_{-0.13} \quad ({\rm Belle}) & 
+0.02\pm 0.34\pm 0.05 \quad ({\rm Babar})
\end{array}
\end{equation}

QCD factorization to leading power in $\Lambda/m_b$ has been
demonstrated at ${\cal O}(\alpha_s)$ for the important class
of decays $B\to\pi\pi$, $\pi K$. For $B\to D\pi$ (class I),
where hard spectator interactions are absent, a proof
has been given explicitly at two loops \cite{BBNS} and
to all orders in the framework of soft-collinear effective theory
(SCET) \cite{BPS}. Complete matrix elements are available
at ${\cal O}(\alpha_s)$ (NLO) for $B\to\pi\pi$, $\pi K$,
including electroweak penguins.
Power corrections are presently not calculable in general.
Their impact has to be estimated and included into the error
analysis. Critical issues here are annihilation
contributions and certain corrections
proportional to $m^2_\pi/((m_u+m_d)m_b)$, which is numerically
sizable, even if it is power suppressed.
However, the large variety of channels available will provide us with
important cross checks and arguments based on SU(2) or SU(3)
flavour symmetries can also be of use in further controling
uncertainties.

\section{Radiative decays $B\to V\gamma$}

Factorization in the sense of QCD can also be applied to the
exclusive radiative decays $B\to V\gamma$ ($V=K^*$, $\rho$).
The factorization formula for the operators in the
effective weak Hamiltonian can be written as \cite{BFS,BB}
\begin{equation}\label{fform}
\langle V\gamma(\epsilon)|Q_i|\bar B\rangle =
\Bigl[ F^{B\to V}(0)\, T^I_{i} 
+\int^1_0 d\xi\, dv\, T^{II}_i(\xi,v)\, \Phi_B(\xi)\, \Phi_V(v)\Bigr]
\cdot\epsilon
\end{equation}
where $\epsilon$ is the photon polarization 4-vector.
Here $F^{B\to V}$ is a $B\to V$ transition form factor,
and $\Phi_B$, $\Phi_V$ are leading twist light-cone distribution amplitudes
(LCDA) of the $B$ meson and the vector meson $V$, respectively.
These quantities 
describe the long-distance dynamics of the matrix elements, which
is factorized from the perturbative, short-distance interactions
expressed in the hard-scattering kernels $T^I_{i}$ and $T^{II}_i$.
The QCD factorization formula (\ref{fform}) holds up to
corrections of relative order $\Lambda_{QCD}/m_b$.
Annihilation topologies are power-suppressed, but still calculable
in some cases. The framework of QCD factorization is necessary to
compute exclusive $B\to V\gamma$ decays systematically beyond the
leading logarithmic approximation. Results to next-to-leading order
in QCD, based
on the heavy quark limit $m_b\gg\Lambda_{QCD}$ have been
computed \cite{BFS,BB}
(see also \cite{AP}).

The method defines a systematic,
model-independent framework for $B\to V\gamma$.
An important conceptual aspect of this analysis is the interpretation
of loop contributions with charm and up quarks, which come from
leading operators in the effective weak Hamiltonian.
These effects are calculable in terms of
perturbative hard-scattering functions and universal meson
light-cone distribution amplitudes. They are ${\cal O}(\alpha_s)$
corrections, but are leading power contributions in the
framework of QCD factorization. This picture is in contrast to the
common notion that considers charm and up-quark loop effects as
generic, uncalculable long-distance contributions.
Non-factorizable long-distance corrections may still exist, but
they are power-suppressed.
The improved theoretical understanding of $B\to V\gamma$ decays
streng\-thens the motivation for still more detailed
experimental investigations, which will contribute
significantly to our knowledge of the flavour sector.

The uncertainty of the branching fractions is 
currently dominated by the form factors $F_{K^*}$, $F_\rho$. 
A NLO analysis \cite{BB} yields 
(in comparison with the experimental results in brackets)
$B(\bar{B}\to \bar{K}^{*0}\gamma)/10^{-5}=7.1\pm 2.5$  
($4.21\pm 0.29$ \cite{BABAR1})
and $B(B^-\to\rho^-\gamma)/10^{-6}=1.6\pm 0.6$ 
($< 2.3$ \cite{BABAR2}).
Taking the sizable uncertainties into 
account, the results for $B\to K^*\gamma$ are compatible with the 
experimental measurements, 
even though the central theoretical values appear to be somewhat high.
$B(B\to\rho\gamma)$ is a sensitive
measure of CKM quantities such as the angle $\gamma$.

\section{Forward-backward asymmetry zero in $B\to K^* l^+l^-$}

Substantial progress has taken place over the last few years
in understanding the QCD dynamics of exclusive $B$ decays.
The example of the forward-backward asymmetry in $B\to K^* l^+l^-$
nicely illustrates some aspects of these developments.

\begin{figure}[t]
\vspace{8cm}
\includegraphics{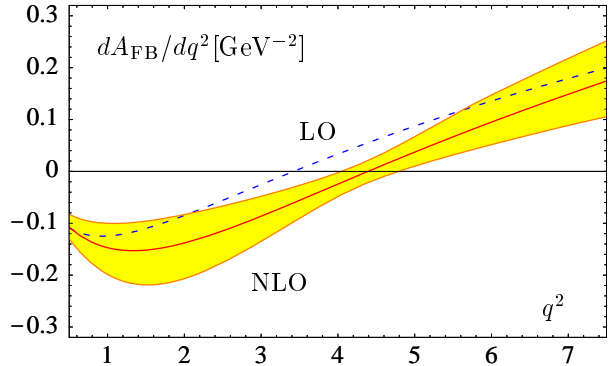}
\vspace{-2cm}
\caption{\it $A_{FB}$ spectrum for $\bar B\to K^*l^+l^-$ at
leading and next-to-leading order in QCD (from \cite{BFS}).
\label{fig:afb}}
\end{figure}

The forward-backward asymmetry $A_{FB}$ is the rate difference
between forward ($0<\theta <\pi/2$) and backward ($\pi/2 <\theta < \pi$)
going $l^+$, normalized by the sum, where $\theta$ is the angle between
the $l^+$ and $B$ momenta in the centre-of-mass frame of the dilepton pair.
$A_{FB}$ is usually considered as a function of the dilepton mass $q^2$.
In the standard model the spectrum $dA_{FB}/dq^2$ (Fig. \ref{fig:afb})
has a characteristic zero at
\begin{equation}\label{afb0}
\frac{q^2_0}{m^2_B}=-\alpha_+ \frac{m_b C_7}{m_B C^{eff}_9}
\end{equation}
depending on short-distance physics contained in the coefficients
$C_7$ and $C^{eff}_9$. The factor $\alpha_+$, on the other hand,
is a hadronic quantity containing ratios of form factors.

It was first stressed in \cite{BUR} that $\alpha_+$ is
not very much affected by hadronic uncertainties and very
similar in different models for form factors with $\alpha_+\approx 2$.
After relations were found between different heavy-light form factors 
($B\to P$, $V$)
in the heavy-quark limit and at large recoil \cite{CLOPR},
it was pointed out in \cite{ABHH} that as a consequence $\alpha_+ =2$ holds  
exactly in this limit. Subsequently, the results of \cite{CLOPR}
were demonstrated to be valid beyond tree level \cite{BFS,BFPS}.
The use of the $A_{FB}$-zero  as a {\it clean\/} test of
standard model flavour physics was thus put on a firm basis and
NLO corrections to (\ref{afb0}) could be computed \cite{BFS}.
More recently also the problem of power corrections to heavy-light form
factors at large recoil in the heavy-quark limit has been studied
\cite{BCDF}.
Besides the value of $q^2_0$, also the sign of the slope
of $dA_{FB}(\bar B)/dq^2$ can be used as a probe of new physics.
For a $\bar B$ meson, this slope is predicted to be positive
in the standard model \cite{BHI}. 

\section{Radiative leptonic decay $B\to l\nu\gamma$}

\begin{figure}[t]
\vspace{4cm}
\includegraphics{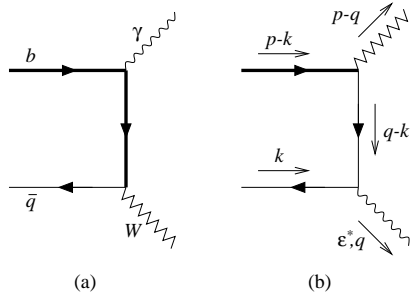}
\vspace{0cm}
\caption{\it Tree-level diagrams for $B\to l\nu\gamma$.
Only diagram (b) contributes at leading power
(see \cite{DGS}).
\label{fig:blnug}}
\end{figure}

The tree-level process $B\to l\nu\gamma$ is not so much of
direct interest for flavour physics, but it provides us with an 
important laboratory for studying QCD dynamics in exclusive
$B$ decays, which is crucial for many other applications.
The leading-power contribution comes from the diagram
in Fig. \ref{fig:blnug} (b), 
which contains a light-quark
propagator that is off-shell by an amount $(q-k)^2\sim q_- k_+$
Here $q$ is the hard, light-like momentum of the photon
with components scaling as $m_b$ (this restricts the region of phase-space
where the present discussion applies), and $k$ is the soft momentum of
the spectator quark. The decay is thus determined by a hard-scattering
process, but also depends on the structure of the $B$ meson in a
non-trivial way \cite{KPY}. 
Recently, in \cite{DGS} it has been proposed, and shown to one loop in QCD, 
that the form factors $F$ for this decay factorize as
\begin{equation}
F=\int d\tilde k_+ \Phi_B(\tilde k_+)\, T(\tilde k_+)
\end{equation}
where $T$ is the hard-scattering kernel and $\Phi_B$ the light-cone
distribution amplitude of the $B$ meson defined as
\begin{equation}
\Phi_B(\tilde k_+)=\int dz_- e^{i\tilde k_+ z_-}
\langle 0|b(0) \bar u(z)|B\rangle|_{z_+=z_\perp=0}
\end{equation}
The hard process is characterized by a scale $\mu_F\sim \sqrt{m_b\Lambda}$.
At lowest order the form factors are proportional to
$\int d\tilde k_+\, \Phi_B(\tilde k_+)/\tilde k_+\equiv 1/\lambda_B$,
a parameter that enters hard-spectator processes in many other
applications. The analysis at NLO requires resummation of
large logarithms $\ln(m_b/\tilde k_+)$.  
An extension of the proof of factorization to all orders was subsequently
given by \cite{LPW} within the SCET.

\section{Conclusions}

Factorization formulas in the heavy-quark limit have been
proposed for a large variety of exclusive $B$ decays.
They justify in many cases the phenomenological factorization
ansatz that has been employed in many applications.
In addition they enable consistent and systematic calculations
of corrections in powers of $\alpha_s$. Non-factorizable
long-distance effects are not calculable in general but they are
suppressed by powers of $\Lambda_{QCD}/m_b$. So far,
$B\to D^+\pi^-$ decays are probably understood best. Decays
with only light hadrons in the final state such as $B\to\pi\pi$,
$K^*\gamma$, $\rho\gamma$, or $K^*l^+l^-$ include hard spectator
interactions at leading power and are therefore more complicated.
An important new tool that has been developed is the soft-collinear
effective theory (SCET), which is of use for proofs of factorization
and for the theory of heavy-to-light form factors at large recoil.
Recent studies of the prototype process $B\to l\nu\gamma$ have
also led to a better understanding of QCD dynamics in exclusive
hadronic $B$ decays.
All these are promising steps towards achieving a good
theoretical control over QCD dynamics in rare hadronic $B$ decays,
which is necessary for probing CP violation, flavour physics
and new phenomena at short distances.

\section*{Acknowledgements}

I thank the organizers of the Frontier Science Workshop
in Frascati for the invitation to a very interesting
and pleasant meeting.

\end{document}